\documentstyle[multicol,aps,psfig,epsfig]{revtex}
\begin{document} 
\draft
\title{Spin-wave spectrum in $\rm La_{2}Cu O_4$ \\
--- double occupancy and competing interaction effects}
\author{Avinash Singh\cite{avinash} and Pallab Goswami}
\address{Department of Physics, 
Indian Institute of Technology Kanpur - 208016, India}
\maketitle
\begin{abstract} 
The recently observed spin-wave energy dispersion along the AF zone boundary 
in $\rm La_{2}Cu O_4$
is discussed in terms of double occupancy and competing interaction effects
in the $t-t'$ Hubbard model on a square lattice.
\end{abstract}
\pacs{PACS nnumbers: 71.10.Fd, 75.10.Lp, 75.30.Ds}  
\begin{multicols}{2}\narrowtext
\section{Introduction}
Recently high resolution inelastic neutron scattering
studies of the spin-wave spectrum 
have been carried out in the square-lattice spin-1/2
antiferromagnet $\rm La_{2}Cu O_4$.\cite{spinwave}
Spin-wave energies were obtained along high symmetry directions
in the two-dimensional (2D) Brillouin zone.
With wavevectors identified by their coordinates $(h,k)$ in the
2D reciprocal space of the square lattice,
these studies reveal a noticeable spin-wave dispersion
along the AF zone boundary,
with a maximum of 335 meV
near the point X having ${\bf Q}=(1/2,0)$
and a minimum of 285 meV near ${\bf Q}=(3/4,1/4)$
at $T=10$ K.

The spin-wave spectrum for the Quantum Heisenberg antiferromagnet
(QHAF) with only nearest-neighbour (NN) exchange coupling $J$,
conventionally obtained using the linear spin-wave theory,
shows no dispersion along the AF zone boundary.
The simplest explanation for the observed spin-wave dispersion 
involves a next-nearest-neighbour (NNN) {\em ferromagnetic}
exchange coupling $J'$ between Cu spins,
and best fit was found with $J=104.1 \pm 4$ meV and
$J'= - 18 \pm 3$ meV at $T=10$ K.\cite{spinwave}

It was pointed out\cite{spinwave}
that a more natural explanation of the data
can be obtained in  terms of the one-band Hubbard model,
for which the strong coupling expansion in powers of the 
NN hopping term $t$ up to fourth order O($t^4$) yields
an effective spin Hamiltonian with {\em antiferromagnetic}
exchange interactions between NN, NNN, and NNNN pairs of spins,
as well as a cyclic ring exchange interaction coupling four spins at
the corners of a square plaquette.\cite{expn1,expn2,expn3} 
The quantitative description within linear spin-wave theory
of both the spin-wave energies and intensities
was taken as a strong demonstration
that the one-band Hubbard model is an excellent starting point
for describing the magnetic interactions in cuprates.
Fits were found to be indistinguishable 
from those for variables $J$ and $J'$,
and yielded $t=0.30 \pm 0.02$ eV and $U=2.2 \pm 0.4$ eV at $T=10$ K,
so that the characteristic ratio $U/t=7.3$. 
Recently, excellent fit has also been obtained 
with the spin-wave energy for the square-lattice Hubbard model, 
evaluated numerically in the Random Phase Approximation (RPA).\cite{num_rpa} 

The one-band Hubbard model with only NN hopping 
possesses particle-hole symmetry, 
implying identical behaviour for hole and electron doping. 
However, there exist significant differences 
in the magnetic properties of hole-doped and electron-doped cuprates. 
In particular, AF order persists in the electron-doped cuprate 
$\rm Nd_{2-x}Ce_x Cu O_4$ up to a doping concentration of
about 15\%,\cite{electron1,electron2}
whereas only 2\% hole concentration destroys AF order 
in $\rm La_{2-x}Sr_x Cu O_4$.
Thus, the Hubbard model with only NN hopping is unable to describe 
the magnetic properties of doped cuprates.  

Recently, the magnetic phase diagram of the $t-t'$ Hubbard model 
has been obtained in the $t'-U$ space,\cite{doped} 
for electron and hole doping and both signs of the NNN hopping term $t'$. 
The phase diagram shows that for an appropriate sign of $t'$,
the AF state survives electron doping 
(up to $x \approx 20\%$ for $|t'/t| = 0.25$ and $U/t \approx 8$),
whereas for any finite hole doping the AF state is destroyed.
Thus the $t-t'$ Hubbard model appears to be the simplest correlated 
electron model capable of describing the magnetic correlations in cuprates, even at the doped level.

The need for more realistic microscopic models,
which include NNN hopping etc., 
has also been acknowledged recently from 
band structure studies, photoemission data
and neutron-scattering measurements of high-T$_{\rm c}$
and related materials.\cite{nnn1,nnn2,nnn3,nnn4}
Estimates for $|t'/t|$ range from 0.15 to 0.5.

Now, the NNN hopping term $t'$ generates 
antiferromagnetic NNN interaction of comparable magnitude 
($t^{'2}/t^2$ is of the order of $J'/J$).
The resulting competition between the exchange interactions
will modify the values of $t$ and $U$ obtained earlier.\cite{spinwave}
In this brief report we obtain the modified one-band Hubbard model parameters by fitting the spin-wave spectrum for $\rm La_{2}Cu O_4$ 
with that for the $t-t'$ Hubbard model.
We have used the RPA approach 
to directly obtain the spin-wave propagator in the AF state
of the $t-t'$ Hubbard model. 
This approach has the advantage 
of being applicable in the intermediate- and weak-coupling regimes as well, 
as it eliminates the need for going through
the strong coupling expansion to obtain the effective spin model.

Finite-$U$ effects of double occupancy on the spin-wave dispersion 
in the half-filled Hubbard model on a square lattice 
have been studied earlier at the RPA level,
both analytically by systematically expanding in powers of $t/U$,\cite{inter} 
and also numerically through an exact computation.\cite{weak} 
This latter approach, desirable in view of the slow, asymptotic
convergence of the $t/U$ expansion,
was employed to study the spin-wave properties in the
weak coupling limit, 
where it was found that the spin-wave energy scale gets 
compressed by the decreasing charge gap.

In ref. [13], the spin-wave propagator at the RPA level,
$\chi^{-+}({\bf q},\omega)= 
\chi^0({\bf q},\omega)/1-U\chi^0({\bf q},\omega)$,
was obtained by keeping all terms up to O($t^4/U^4$) 
(and some terms even up to O($t^6/U^6$))
in the zeroth-order particle-hole propagator $\chi^0({\bf q},\omega)$. 
The terms causing spin-wave dispersion along the 
AF zone boundary are of order $t^2/U^2$,
showing the importance of charge fluctuations,
as the double occupancy term 
$\langle n_{i\uparrow} \, n_{i\downarrow}\rangle$
is of the same order. 
This RPA approach is isomorphous  
to the linear spin-wave analysis\cite{expn3,lsw2}
of the effective spin-1/2 Heisenberg model obtained from
the strong-coupling expansion up to order $t^4$,
where the higher order exchange terms arise from the coherent motion of
electrons beyond the NN sites.\cite{expn1,expn2,expn3}

\section{Spin-wave spectrum}
Including the contribution in $\chi^0({\bf q},\omega)$ of the NNN hopping term $t'$
up to the leading order ($t^{'2}/U^2$) only,\cite{phase} 
and neglecting terms of order $t^2 t^{'2}/U^4$, and $t^{'4}/U^4$ 
which are an order of magnitude smaller than $t^4/U^4$, 
we obtain the spin-wave propagator
\begin{eqnarray}
& & \chi^{-+}({\bf q},\omega) = \nonumber \\
& - &\frac{1}{2} M \left [\begin{array}{lr}
a-\frac{\omega}{2J}  & -b\gamma_{\bf q}  \\
-b\gamma_{\bf q}   & a+\frac{\omega}{2J}   
\end{array}\right ] 
\times \frac{2J}{\omega_{\bf q}}
\left (\frac{1}{\omega-\omega_{\bf q}} - 
\frac{1}{\omega+\omega_{\bf q}}
\right ) \; , \nonumber \\
\end{eqnarray}
and the spin-wave energy 
\begin{equation}
\left ( \frac{\omega_{\bf q}}{2J} \right )^2 =
(1-\gamma_{\bf q} ^2 ) -
\frac{4t^2}{U^2} (6+3\gamma'_{\bf q} -9 \gamma_{\bf q} ^2) 
- \frac{2t^{'2}}{t^2}(1-\gamma'_{\bf q}) \; ,
\end{equation}
where $\gamma_{\bf q} = (\cos q_x + \cos q_y)/2$ 
and $\gamma ' _{\bf q} = \cos q_x \cos q_y$,
and $M=1-8t^{2}/U^2$ is the HF-level sublattice magnetization.
Here ${\bf q}={\bf Q} - (\pi,\pi)$ is the wavevector measured
from the zone center. 
The exchange energy scale $J=4t^2/U$, 
and the matrix elements $a$ and $b$ in Eq. (1) are given by  
\begin{eqnarray}
 a&=&1-\frac{4t^{2}}{U^2}\left (3+\frac{3}{2}\gamma ' _{\bf q} 
+\gamma_{\bf q}^{2}\right ) -\frac{t^{'2}}{t^2}(1-\gamma'_{\bf q}) \nonumber \\
b&=&1-\frac{4t^{2}}{U^2}\left (\frac{11}{2}\right ) \; ,
\end{eqnarray}
The form of Eq. (1) ensures that the spin susceptibility sum rule, 
$\int\frac{d\omega}{2\pi i}[\chi^{+-}(\omega)-\chi^{-+}(\omega)]
=M\sigma_{z}$,  is obeyed.\cite{forster}
This feature of the spin-wave propagator was also noted when 
quantum corrections were included,\cite{quantum} 
where it is the one-loop-level sublattice magnetization,
reduced by quantum spin fluctuations, which appears as the overall factor.

\begin{figure}
\vspace*{-80mm}
\hspace*{-40mm}
\psfig{file=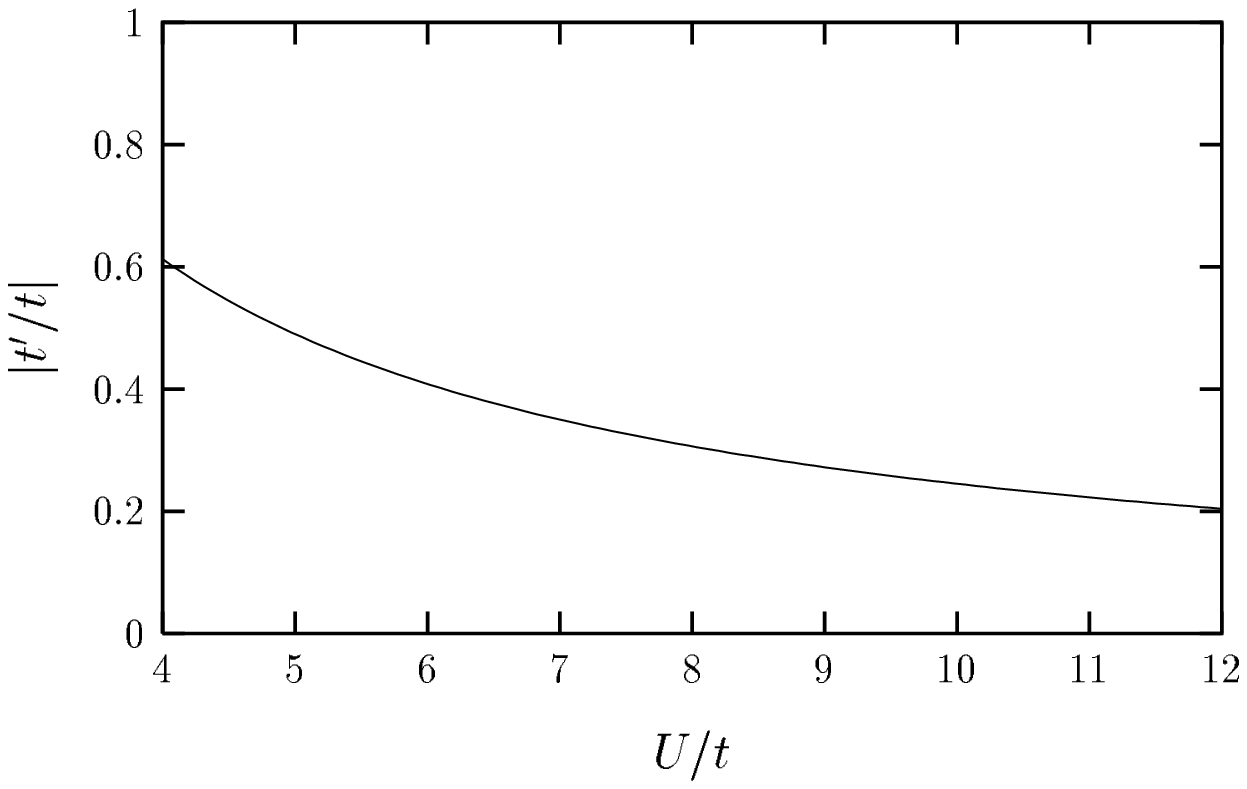,width=145mm,height=220mm,angle=0}
\vspace{-80mm}
\caption{The $|t'/t|$ value below which $\omega(\pi,0) > \omega(\pi/2,\pi/2)$, 
as seen for $\rm La_{2}Cu O_4$ in the neutron-scattering studies.}
\end{figure}

In Eq. (2) for the spin-wave energy, 
the double-occupancy term of order $t^2/U^2$ 
and the competing-interaction term of order $t^{'2}/t^2$ 
are of similar magnitude.
While both terms remove the degeneracy in the spin-wave spectrum 
along the AF zone boundary $(\gamma_{\bf q}=0)$,
they give rise to {\em opposite} dispersion along the AF zone boundary.
Thus, while double occupancy leads to maximum and minimum 
spin-wave energies 
at points ${\bf q}=(\pm \pi,0),(0,\pm \pi)$ and $(\pm \pi/2,\pm \pi/2)$
respectively, the competing interaction favors an exactly opposite trend. 

\ \\
\begin{center}
{\small \bf III. DETERMINATION OF $t$ and $U$} \\
\end{center}

From Eq. (2), the spin-wave velocity 
$c=\lim_{q\rightarrow 0} (\omega_{\bf q}/q )$, 
and the spin-wave energies at points $(\pi,0)$ and $(\pi/2,\pi/2)$ are obtained as:
\begin{eqnarray}
c & = & \sqrt{2} J Z_c \left (1-\frac{24t^2}{U^2} - \frac{2t^{'2}}{t^2} \right )^{1/2} \\
\omega(\pi,0)
& = & 2J Z_c \left (1-\frac{12t^2}{U^2} - \frac{4t^{'2}}{t^2} \right )^{1/2} \\
\omega(\pi/2,\pi/2)
& = & 2J Z_c \left (1-\frac{24t^2}{U^2} - \frac{2t^{'2}}{t^2} \right )^{1/2} 
\end{eqnarray}
A momentum-independent multiplicative renormalization 
factor $Z_c =1.18$,
arising from quantum spin fluctuations,\cite{zc} 
has been included.
Equations (4-6) clearly show that for $U/t \sim 8$ and $t'/t \sim 1/4$
(as appropriate for the cuprates),
the double occupancy and competing interaction terms 
are of the same order of magnitude.
It is also interesting to note that $c$ and $\omega(\pi/2,\pi/2)$ 
are directly proportional to each other. Indeed, the 
spin-wave spectrum for $\rm La_{2}Cu O_4$ shows  
that both $c$ and $\omega(\pi/2,\pi/2)$
are nearly unchanged with temperature.\cite{spinwave} 
Interestingly, from Eqs. (4) and (6) we find that 
$c$ and $\omega(\pi/2,\pi/2)$ have a vanishing
derivative with respect to $t/U$ 
(which should change with temperature) at $U/t \sim 6$,
which is very close to the $U/t$ value obtained 
from our fit.  

As the maximum in the spin-wave energy spectrum of $\rm La_{2}Cu O_4$  
is clearly seen to occur at $(\pi,0)$, 
the double occupancy term is actually dominant,
showing the importance of charge fluctuations in cuprates.
From this information, an upper limit for the ratio $|t'/t|$ can be deduced
by comparing the spin-wave energies $\omega(\pi,0)$ and 
$\omega(\pi/2,\pi/2)$. 
From Eqs. (5) and (6), we obtain the condition 
$t^{'2}/t^2 < 6t^2/U^2$, which is shown in Fig. 1.
For $U/t \sim 8$, the ratio $|t'/t|$ 
must be smaller than $\sim 0.3$.

Taking the key values of the 
maximum spin-wave energy $\omega(\pi,0)$ 
and the spin-wave velocity $c$ from the spin-wave spectrum,  
we solve for $t/U$ and $J$ from Eqs. (4) and (5),
and thus obtain $t$ and $U$.     
With $\omega(\pi,0) = 335$ meV and $c = 200$ meV at $T=10$ K, 
and taking $|t'/t|=0.25$, 
we obtain $t^2/U^2 = .022$, $U/t=6.7$, $t=0.34$ eV, and $U=2.3$ eV.
Figure 2 shows the spin-wave spectrum calculated from Eq. (2) with these
$t$ and $U$ values,
and the experimental points for $\rm La_{2}Cu O_4$ at 10 K. 
Without $t'$, the corresponding values are
$t^2/U^2 = .019$, $U/t=7.3$, $t=0.29$ eV, and $U=2.2$ eV
in agreement with the results of ref. [1], and showing a 
nearly $20\%$ enhancement in the double occupancy
factor $t^2/U^2$ due to the competing effect of $t'$.
With $\omega(\pi,0) = 320$ meV and $c = 200$ meV
at $T=295$ K, we obtain $U/t=7.1$, $t=0.34$ eV, and $U=2.4$ eV.

In the approach adopted in ref. [1],
the effective spin $S=1/2$ quantum Heisenberg Hamiltonian
obtained by the strong-coupling expansion of the Hubbard model
contains all physical processes up to the order $t^4$ level,
and the subsequent use of the linear spin-wave approximation 
to obtain the spin-wave spectrum neglects the quantum corrections
of order $1/S$.
In our approach these two steps are carried out in exactly reverse order.
The evaluation of the spin-wave propagator in the random phase
approximation neglects the quantum corrections 
(of equivalent order $1/{\cal N} = 1/2S$ within the inverse-degeneracy expansion
of the generalized ${\cal N}$-orbital Hubbard model)\cite{quantum}.
The evaluation of $\chi^0({\bf q},\omega)$ is next carried out,
either analytically via a systematic expansion in powers of $t/U$,
or numerically through an exact computation.

The spin-wave dispersion along the AF zone boundary, 
caused by the double occupancy effect at finite $U$
and partially neutralized by the competing interaction effect of $t'$,
results in a broadening of the spin-wave spectrum 
at the high-energy end, 
resembling a spin-wave damping effect. 
Implication of this broadening to
the two-magnon Raman scattering in 
$\rm La_{2}Cu O_4$\cite{raman1,raman2}
has been discussed earlier.\cite{inter,raman}
The importance of charge fluctuation effects 
(in terms of the cyclic exchange terms)
have also been pointed out 
in infrared absorption experiments,\cite{ir1,ir2}
and magnetic properties of the related compound
$\rm Sr_{14} Cu_{24} O_{41}$.\cite{related}

\begin{figure}
\vspace*{0mm}
\hspace*{-10mm}
\psfig{file=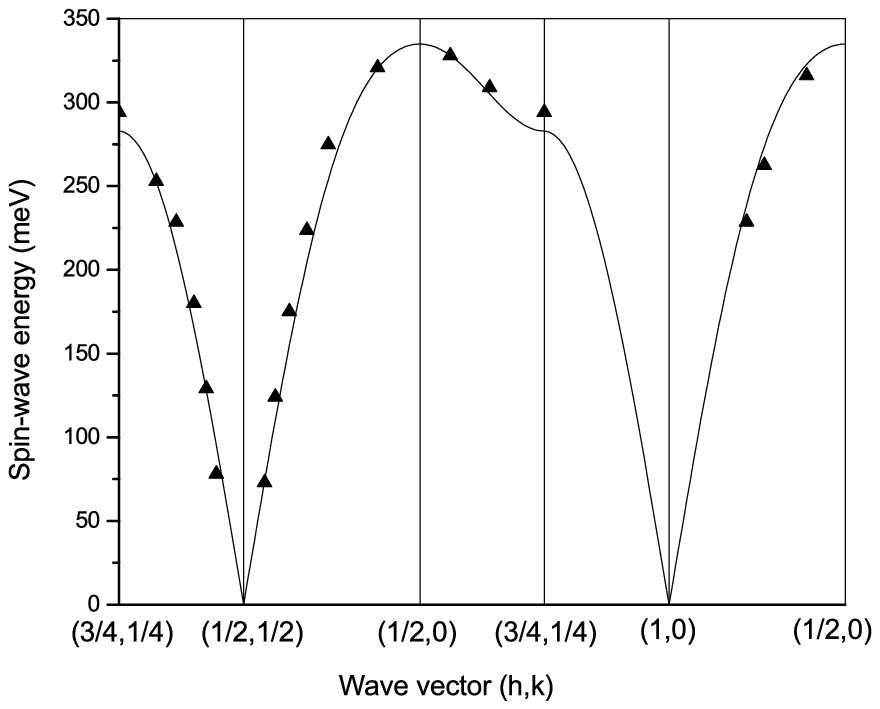,height=60mm,width=95mm,angle=0}
\vspace{0mm}
\caption{The calculated spin-wave energy spectrum $\omega_{\bf q}$ 
from Eq. (2) (line) and spin-wave energies for $\rm La_{2}Cu O_4$ at $T=10$ K 
(triangles) from ref. [1].}
\end{figure}

In conclusion, the competing interaction effect of the 
NNN hopping term $t'$,
which renders the Hubbard model more realistic
for describing the magnetic properties of the doped cuprates,
partially neutralizes the spin-wave dispersion 
along the AF zone boundary caused by the double occupancy effect.  
Due to this competition the double occupancy factor $t^2/U^2$ 
must be significantly enhanced by nearly $20\%$
in order to fit the observed spin-wave spectrum 
in $\rm La_{2}Cu O_4$.

\end{multicols}

\begin{references}

\bibitem[\dag]{avinash}
Electronic address: avinas@iitk.ac.in

\bibitem[1]{spinwave}
R. Coldea, S. M. Hayden, G. Aeppli, T. G. Perring, C. D. Frost,
T. E. Mason, S.-W.Cheong, and Z. Fisk,
Phys. Rev. Lett. {\bf 86}, 5377 (2001).

\bibitem[2]{expn1}
M. Takahashi, J. Phys. C {\bf 10}, 1289 (1977).

\bibitem[3]{expn2}
M. Roger and J. M. Delrieu, Phys. Rev. B {\bf 2299}, (1989).

\bibitem[4]{expn3}
A. H. MacDonald, S. M. Girvin, and D. Yoshioka, 
Phys. Rev. B {\bf 41}, 2565 (1990); {\bf 37}, 9753 (1988).

\bibitem[5]{num_rpa}
N. M. R. Peres and M. A. N. Ara\'{u}jo,
Phys. Rev. B {\bf 65}, 132404 (2002).

\bibitem[6]{electron1}
C. Almason and M. B. Maple, 
in {\em Chemistry of High-Temperature Superconductors},
edited by C. N. R. Rao, World Scientific, Singapore (1991).

\bibitem[7]{electron2}
M. Matsuda, Y. Endoh, K. Yamada, H. Kojima, I. Tanaka,
R. J. Birgeneau, M. A. Kastner, and G. Shirane,
Phys. rev. {\bf 45}, 12548 (1992).

\bibitem[8]{doped}
A. Singh and H. Ghosh, 
Phys. Rev. B {\bf 65}, 134414 (2002). 

\bibitem[9]{nnn1}
P. B\'{e}nard, L. Chen, and A. -M. S. Tremblay,
Phys. Rev. B {\bf 47}, 589 (1993); \\
A. Veilleux, A. Dar\'{e}, L. Chen,
Y. M. Vilk, and A. -M. S. Tremblay, Phys. Rev. B {\bf 52}, 16255 (1995).

\bibitem[10]{nnn2}
T. Tohyama and S. Maekawa, Phys. Rev. B {\bf 49}, 3596 (1993).

\bibitem[11]{nnn3}
G. Stemmann, C. P\'{e}pin, and M. Lavagna, Phys. Rev. B
{\bf 50}, 4075 (1994).

\bibitem[12]{nnn4}
O. K. Andersen, A. I. Liechtenstein, O. Jepsen, and F. Paulsen,
J. Phys. Chem. Solids {\bf 56}, 1573 (1995).

\bibitem[13]{inter}
A. Singh, Phys. Rev. B {\bf 48}, 6668 (1993).

\bibitem[14]{weak}
P. Sen and A. Singh, 
Phys. Rev. B {\bf 48}, 15792 (1993).

\bibitem[15]{lsw2}
A. Chubukov, E. Gagliano, and C. Balseiro,
Phys. Rev. B {\bf 45}, 7889 (1992).

\bibitem[16]{phase}
A. Singh, 
Report No. cond-mat/0112442 (2001). 

\bibitem[17]{forster}
Dieter Forster, {\em Hydrodynamic Fluctuations, Broken Symmetry, and 
Correlation Functions}, Frontiers in Physics Lecture Note Series Vol. 47 
(Benjamin/Cummings, New York, 1975), p. 198.

\bibitem[18]{quantum}
A. Singh, Phys. Rev. B {\bf 43}, 3617 (1991).

\bibitem[19]{zc}
R. R. P. Singh, Phys. Rev. B {\bf 39}, 9760 (1989).

\bibitem[20]{raman1}
K. B. Lyons {\em et al.},  
Phys. Rev. B {\bf 39}, 9693 (1989).

\bibitem[21]{raman2}
S. Sugai {\em et al.},  
Phys. Rev. B {\bf 42}, 1045 (1990).

\bibitem[22]{raman}
S. Basu and A. Singh, 
Phys. Rev. B {\bf 54}, 6356 (1996). 

\bibitem[23]{ir1}
J. Lorenzana, J. Eroles, and S. Sorella,
Phys. Rev. Lett. {\bf 83}, 5122 (1999).

\bibitem[24]{ir2}
J. D. Perkins {\em et al.}, 
Phys. Rev. Lett. {\bf 71}, 1621 (1993).

\bibitem[25]{related}
S. Brehmer {\rm et al.}, 
Phys. Rev. B {\bf 60}, 329 (1999). 

\end{references}
\end{document}